\documentclass[a4paper,12pt]{article}

\usepackage{a4}
\usepackage{graphics}

\begin{document}
\begin{center}
{\Large \bf Possible Test of the GUT Relation between $M_1$ and $M_2$ in Electron-Photon Scattering}
\end{center}
\vspace{0.5cm}
\begin{center}
{\large \sf Claus Bl\"ochinger\footnote{e-mail: bloechi@physik.uni-wuerzburg.de}, Hans Fraas\footnote{e-mail: fraas@physik.uni-wuerzburg.de}}
\end{center}
\begin{center}
{\small \it Institut f\"ur Theoretische Physik, Universit\"at W\"urzburg, Am 
Hubland, \\
D-97074 W\"urzburg, Germany}
\end{center}
\begin{abstract}
We investigate associated production  of selectrons and the lightest 
neutralino (LSP) in the process 
$e^-\gamma \longrightarrow \tilde{\chi}_1^0\tilde{e}_{L/R}^-$ with 
the selectron subsequently decaying into an electron and the LSP.
Total cross sections and various polarization asymmetries are calculated for 
photons produced by Compton back\-scattering of a polarized laser beam at an 
$e^+e^-$ linear collider with CMS energy $\sqrt{s_{ee}}=500$ GeV and with 
polarized beams. The total cross section and in particular the polarization 
asymmetries show a characteristic dependence on the gaugino mass parameter 
$M_1$. Therefore this process is suitable for testing the GUT relation 
$M_1=M_2\cdot\frac{5}{3}\tan^2\theta_W $.
\end{abstract}

\section{Introduction}
The search for supersymmetry (SUSY) \cite{susy} is one of the most important 
goals of a future
$e^+e^-$ linear collider (LC) in the energy range between 500 GeV and 1000 GeV \cite{LC}.  In addition to the $e^+e^-$ option the 
$e^-\gamma$ mode is also technically realizable with high luminosity polarized photon
 beams obtained by backscattering of intensive laser pulses off the electron beam \cite{PLCX,PLCY,PLC1}. Associated production  of selectrons with the 
lightest neutralino $\tilde{\chi}_1^0$
(assumed to be the LSP) in $e^-\gamma$ collisions allows to  probe 
heavy selectrons beyond the kinematical limit of selectron pair production in
$e^+e^-$ annihilation. Further
associated production of selectrons and gaugino-like neutralinos provides us with the 
possibility to study the 
electron-selectron-neutralino couplings complementary to $e^+e^-$ annihilation.

In the present paper we study the associated production $e^-\gamma \longrightarrow \tilde{\chi}_1^0 \tilde{e}_{L/R}^-$ with polarized beams and the subsequent direct 
leptonic decay $\tilde{e}_{L/R}^-\longrightarrow 
\tilde{\chi}_1^0e^-$. The beam polarization is chosen suitably  to optimize cross
sections and polarization asymmetries. The signal is a single electron with high 
transverse
momentum $p_T$. We do not consider cascade decays of heavy selectrons, which 
may yield a similar single electron signal with, however, a less pronounced 
$p_T$ \cite{PLC1}. We also refrain from a discussion of the background. 

The calculations are done in the Minimal Supersymmetric Standard Model \linebreak[4](MSSM). The masses and couplings of the neutralinos depend
 on the gaugino mass parameters $M_1$ and $M_2$, the higgsino mass parameter $\mu$ and the ratio $\tan\beta$ of the two Higgs vacuum expectation values. 
The parameters $M_2$, $\mu$ and $\tan\beta$ can in principle be determined by 
chargino production alone \cite{otherM1}.
For the gaugino mass 
parameters usually the GUT relation $M_1=M_2\cdot\frac{5}{3}\tan^2\theta_W$ is
assumed. 
A precise determination of $M_1$ is, however, only possible in the neutralino
sector \cite{neutrsec}.

In the present paper we investigate if associated production of selectrons and the LSP $\tilde{\chi}_1^0$ is suitable as a test for this relation. We therefore study the influence of the gaugino mass 
parameter $M_1$ on the total cross section and on polarization asymmetries for different selectron masses.

\section{Cross Sections and Polarization Asymmetries}

The production cross section $\sigma_P^{L/R}\left(s_{e\gamma}\right)$ for the
process 
$e^-\gamma \longrightarrow \tilde{\chi}_1^0\tilde{e}_{L/R}^-$ proceeds via electron 
exchange in the 
s-channel and
  selectron exchange in the t-channel.
The electron-selectron-LSP 
couplings 
\begin{equation}\label{coupl}
f_{e1}^L=-\sqrt{2}\left[\frac{1}{\cos\theta_W}\left(-\frac{1}{2}+\sin^2\theta_W\right)N_{12}-\sin\theta_WN_{11}\right],
\end{equation}
\begin{equation}\label{coupr}
f_{e1}^R=\sqrt{2}\sin\theta_W\left[\tan\theta_WN_{12}^*-N_{11}^*\right]
\end{equation}
for left and right selectrons with masses $m_{\tilde{e}_L}$ and
$m_{\tilde{e}_R}$ depend on the photino component
 $N_{11}$ and the zino component $N_{12}$ of the LSP \cite{susy}. 
For an electron beam with longitudinal polarization $P_e$ the cross sections 
$\sigma_P^L$ and $\sigma_P^R$ are proportional to $\left(1-P_e\right)$ and 
$\left(1+P_e\right)$, respectively.
For special cases the cross sections are given in \cite{PLC1} and \cite{Grifhols}, the
complete analytical expressions for the differential and the total cross section for 
polarized beams 
will be given in a forthcoming paper \cite{forthcoming}.

In the narrow width approximation one obtains the total cross section 
$\sigma_{e\gamma}^{L/R}$ for the combined process of $\tilde{e}_{L/R}^-$$\tilde{\chi}_1^0$ production and the subsequent leptonic decay $\tilde{e}_{L/R}^-\longrightarrow e^-\tilde{\chi}_1^0$ by multiplying the production cross section with the leptonic branching ratio:
\begin{equation}
\sigma_{e\gamma}^{L/R}\left(s_{e\gamma}\right) = 
\sigma_P^{L/R}\left(s_{e\gamma}\right) \cdot {\rm Br}\left(\tilde{e}_{L/R}^-\longrightarrow e^-\tilde{\chi}_1^0 \right).
\end{equation} 
 
The LSP-selectron-electron coupling  $f_{e1}^{L/R}$
appears in the production amplitudes as well as in the decay amplitude, so
that the total cross section $\sigma_{e\gamma}^{L/R}\left(s_{e\gamma}\right)$ is proportional to $\left(f_{e1}^{L/R}\right)^4$.
 
The photon beam is assumed to be produced by Compton backscattering of circularly polarized laser photons (polarization $\lambda_L$) off longitudinally polarized electrons (polarization $\lambda_e$). The energy spectrum $P\left(y\right)$ and the mean helicity $\lambda\left(y\right)$ of the high energy
photons are given in \cite{PLCY,PLC1,energyspectrum}. The 
ratio $y=E_{\gamma}/E_e$ of the photon energy $E_{\gamma}$ and the energy of the 
converted electron beam $E_e$ is confined to $y\stackrel{\textstyle<}{\sim} 0.83$ 
\cite{PLCX}. 
For $y>0.83$ $e^+e^-$ pairs can be produced via scattering of laser photons and 
backscattered photons,  so that the flux of high energetic photons drops considerably.
To obtain the total cross section 
$\sigma_{ee}^{L/R}\left(s_{ee},P_e,\lambda_e,\lambda_L\right)$ for the combined process in the laboratory frame ($e^+e^-$ CMS) one has to convolute the total cross section $\sigma_{e\gamma}^{L/R}\left(s_{e\gamma}\right)$ in the $e\gamma$ CMS 
 with  the energy distribution $P\left(y\right)$ and the mean helicity $\lambda\left(y\right)$ of the  backscattered photon beam \cite{stefan}:
\begin{equation}
\sigma_{ee}^{L/R} = \int dy P\left(y\right)\hat{\sigma}_{e\gamma}^{L/R}\left(s_{e\gamma}=ys_{ee}\right),
\end{equation}
\begin{eqnarray} \label{sigmahat}
\hat{\sigma}_{e\gamma}^{L/R} & = & \frac{1}{2}\left(1+\lambda\left(y\right)\right)\left(\sigma_{e\gamma}^{L/R}\right)^++\frac{1}{2}\left(1-\lambda\left(y\right)\right)\left(\sigma_{e\gamma}^{L/R}\right)^-\nonumber\\
& = & \sigma_{e\gamma}^{L/R}\left(1+\lambda\left(y\right) A_c^{L/R}\right).
\end{eqnarray} 
In eq. (\ref{sigmahat}) $\left(\sigma_{e\gamma}^{L/R}\right)^{+/-}$ are the total 
cross sections for a completely right (left) circular polarized photon beam 
whereas $\sigma_{e\gamma}^{L/R}$ is the cross section for unpolarized photons.
\begin{equation}
A_c^{L/R}=\frac{\left(\sigma_{e\gamma}^{L/R}\right)^+-\left(\sigma_{e\gamma}^{L/R}\right)^-}{\left(\sigma_{e\gamma}^{L/R}\right)^++\left(\sigma_{e\gamma}^{L/R}\right)^-}
\end{equation}
 is the polarization asymmetry for circular polarized photons.

Since the production and decay of right and left selectrons lead to
the same final state  we  add both cross sections and obtain
\begin{equation}
\sigma_{ee}=\sigma_{ee}^L+\sigma_{ee}^R.
\end{equation}

We consider two types of polarization asymmetries of the convoluted
cross section. For the first one we flip the electron polarization 
$P_e$  and fix the polarization $\lambda_L$ of the laser beam and the polarization $\lambda_e$ of the converted electron beam:
\begin{equation}\label{Ae}
A_{P_e}=\frac{\sigma_{ee}\left(s_{ee},P_e,\lambda_e,\lambda_L\right)-\sigma_{ee}\left(s_{ee},-P_e,\lambda_e,\lambda_L\right)}{\sigma_{ee}\left(s_{ee},P_e,\lambda_e,\lambda_L\right)+\sigma_{ee}\left(s_{ee},-P_e,\lambda_e,\lambda_L\right)}.
\end{equation}
If we split off from $\sigma_{ee}^{L/R}$ the dependence of beam polarization 
$\left(1\mp P_e\right)$ 
\begin{equation}
\sigma_{ee}\left(s_{ee},P_e,\lambda_e,\lambda_L
\right)=\left(1-P_e\right)\tilde{\sigma}_{ee}^L+\left(1+P_e\right)\tilde{\sigma}_{ee}^R,
\end{equation}
we obtain 
\begin{equation}\label{Ae2}
A_{P_e}=P_e\cdot\frac{\tilde{\sigma}_{ee}^R-\tilde{\sigma}_{ee}^L}{\tilde{\sigma}_{ee}^R+\tilde{\sigma}_{ee}^L}.
\end{equation} 
Here $\tilde{\sigma}_{ee}^R$ ($\tilde{\sigma}_{ee}^L$) is the cross section for production of right (left) selectrons with an unpolarized electron beam ($P_e=0$)
and their subsequent leptonic decay. 

As a second asymmetry we discuss that with respect to the polarization $\lambda_L$ of the laser beam:
\begin{equation}\label{AL}
A_{\lambda_L}=\frac{\sigma_{ee}\left(s_{ee},P_e,\lambda_e,\lambda_L\right)-\sigma_{ee}\left(s_{ee},P_e,\lambda_e,-\lambda_L\right)}{\sigma_{ee}\left(s_{ee},P_e,\lambda_e,\lambda_L\right)+\sigma_{ee}\left(s_{ee},P_e,\lambda_e,-\lambda_L\right)}.
\end{equation}

\section{Numerical Results}

In the following numerical analysis we study the total cross section 
$\sigma^{(L/R)}_{ee}$ and the 
polarization asymmetries $A_{P_e}$ and $A_{\lambda_L}$ for $\sqrt{s_{ee}}=500$
GeV. For the MSSM para\-meters we choose
$M_2=152$ GeV, $\mu=316$ GeV, $\tan\beta=3$ with $M_1$ varying between
$M_1=40$ GeV and $M_1=300$ GeV. The region $M_1<40$ GeV is excluded by assuming
a lower limit of 35 GeV for the LSP mass $m_{\tilde{\chi}_1^0}$. In the figures the
excluded region is shaded.
For $M_1=78.7$ GeV this corresponds to the 
DESY/ECFA reference scenario for the Linear Collider \cite{szen1}, which 
implies the GUT relation $M_1=M_2\cdot\frac{5}{3}\tan^2\theta_W$. 

\begin{figure}[htb]
\label{cross-section}
\centering
\begin{picture}(14,10.5)
\put(-1.4,1.9){\includegraphics{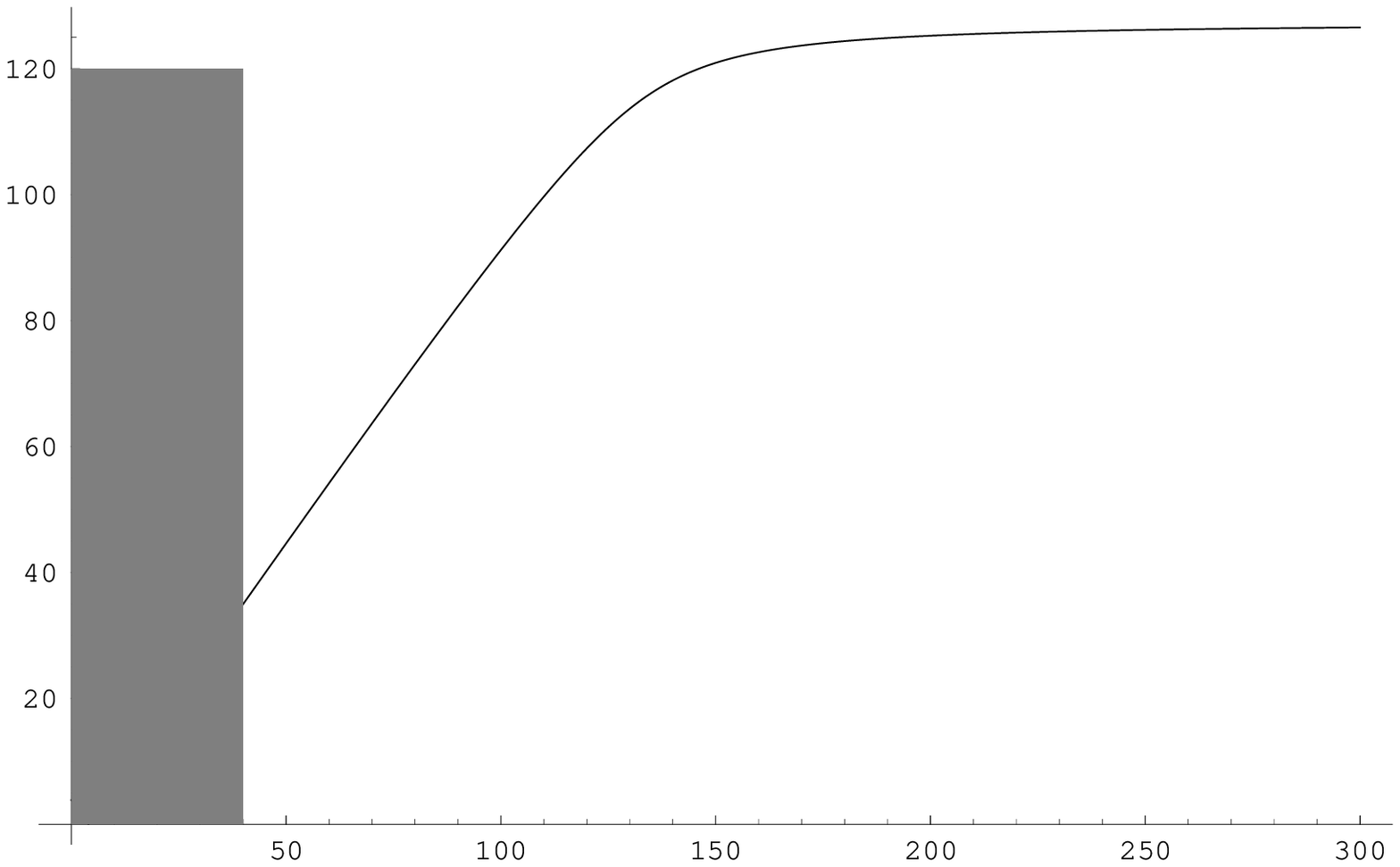}}
\put(5.9,5.25){{\tiny $M_1$/GeV}}
\put(-0.4,9.9){{\tiny $m_{\tilde{\chi}_1^0}$/GeV}}
\put(2.5,5.2){{\tiny (a)}}
\put(6.0,1.9){\includegraphics{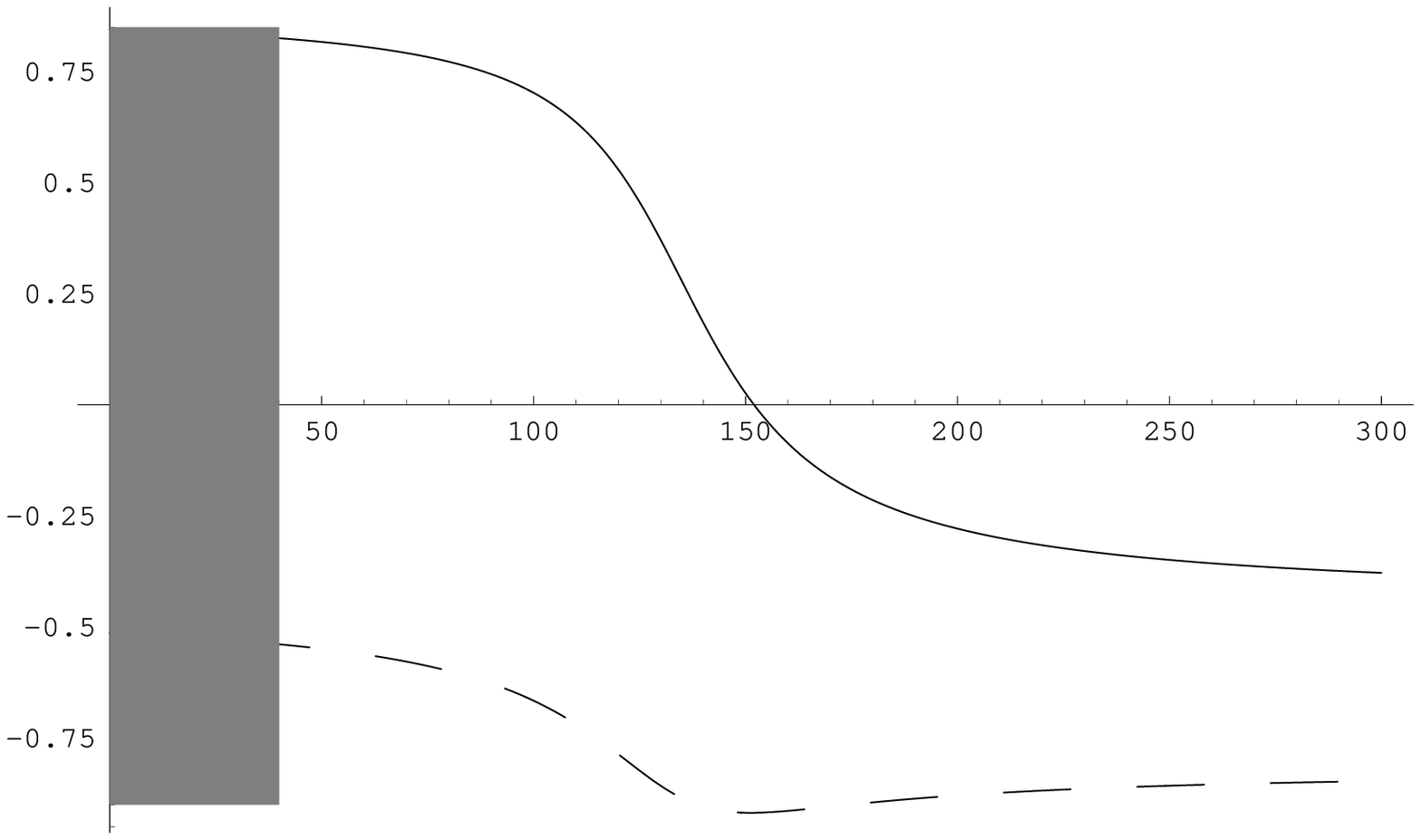}}
\put(13.3,7.8){{\tiny $M_1$/GeV}}
\put(7.0,9.7){{\tiny $N_{11}$,$N_{12}$}}
\put(9.9,5.2){{\tiny (b)}}
\put(2.3,-3.3){\includegraphics{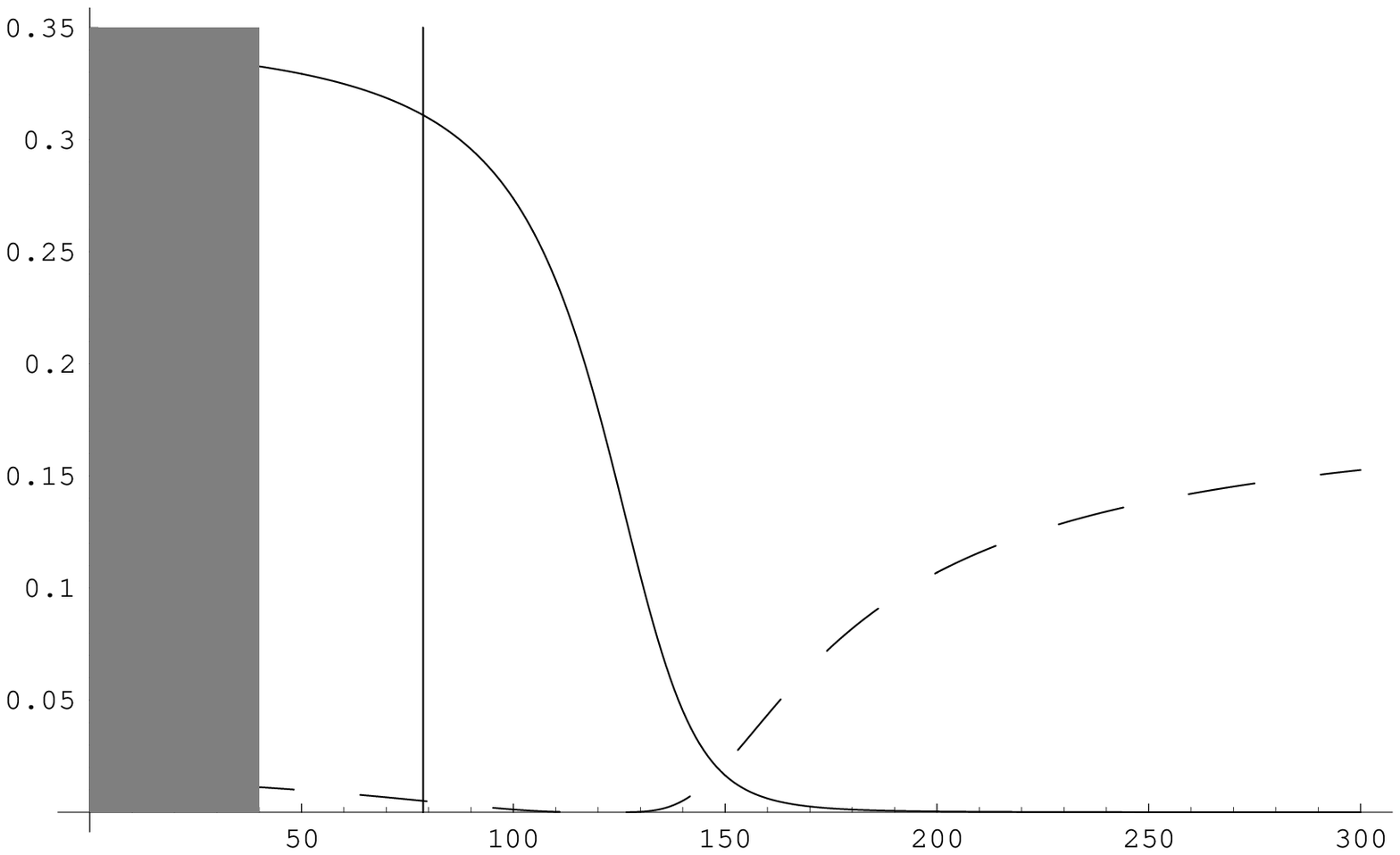}}
\put(9.7,0.05){{\tiny $M_1$/GeV}}
\put(3.3,4.55){{\tiny $\left(f_1^{L/R}\right)^4$}}
\put(6.7,4.0){{\tiny $M_1=78.7$ GeV}}
\put(6.6,4.1){\vector(-1,0){1}}
\put(6.2,0.0){{\tiny (c)}}
\end{picture}
\caption{(a) $M_1$-dependence of the LSP mass $m_{\tilde{\chi}_1^0}$; 
(b) $M_1$-dependence of the photino component $N_{11}$ (solid line) and of 
the zino component $N_{12}$ (dashed line) of the LSP; 
(c) $M_1$-dependence of the couplings $\left(f_{e1}^R\right)^4$ 
(solid line) and $\left(f_{e1}^L\right)^4$ (dashed line).}
\end{figure}

For this set of 
parameters one has 
$35 \mbox{ GeV}<m_{\tilde{\chi}_1^0}<m_{\tilde{\chi}_1^{\pm}}<128$ GeV.
Fig.\ 1a shows that in the region 
$40 \mbox{ GeV}<M_1<150$ GeV the LSP mass depends very strongly on $M_1$, 
varying between 
$m_{\tilde{\chi}_1^0}=35$ GeV for $M_1=40$ GeV and $m_{\tilde{\chi}_1^0}=121$
GeV for $M_1=150$ GeV whereas for $M_1>150$ GeV the mass of the LSP is 
practically 
independent of $M_1$. In the whole $M_1$ region the LSP is 
gaugino-like (fig.\ 1b). At $M_1=M_2$ the photino component $N_{11}$ changes its
sign which leads
to completely different strength of the couplings $f_{e1}^{L/R}$ in the 
regions $M_1>150$ GeV and $M_1<150$ GeV (fig.\ 1c).
For the selectron masses we choose two examples: 
$m_{\tilde{e}_L}=179.3$ GeV, $m_{\tilde{e}_R}=137.7$ GeV corresponding to
the value $m_0=110$ GeV of the common scalar mass at the GUT scale and 
$m_{\tilde{e}_L}=350.0$ GeV, $m_{\tilde{e}_R}=330.5$ GeV corresponding to 
$m_0=320$ GeV. In the second case selectron
pair production at an $e^+e^-$ collider with $\sqrt{s_{ee}}=500$ GeV is
kinematically forbidden.

For the integrated luminosity of the $e\gamma$ machine we assume 
$\int\mathcal{L}=100$ fb$^{-1}$ so that cross sections of a few fb should be
measurable. 

Fig.\ 1c shows that in our scenario also the electron-selectron-LSP couplings 
strongly depend
on $M_1$. For $M_1<150$ GeV the coupling of 
the right selectron $f_{e1}^R$ dominates whereas for $M_1>150$ GeV that of the left 
selectron $f_{e1}^L$ is the stronger one.
 Similarly the 
total cross sections
$\sigma^{L/R}_{ee}$ depicted in fig.\ 2a for a CMS energy $\sqrt{s_{ee}}=500$ 
GeV and for unpolarized beams ($P_e=\lambda_L=\lambda_e=0$) have a 
pronounced $M_1$-dependence. 
Comparing fig.\ 2a  for the cross sections with fig.\ 1c for the couplings
$f_{e1}^{L/R}$ one can see that even in the region $40\mbox{ GeV}<M_1<150$ GeV the 
influence of the additional $M_1$-dependence
of the LSP mass (fig.\ 1a) is weak so that the total cross sections reflect 
essentially the $M_1$-dependence of the couplings.

\begin{figure}[h]
\label{f1LRhoch4}
\centering
\begin{picture}(14,10.5)
\put(-1.4,1.9){\includegraphics{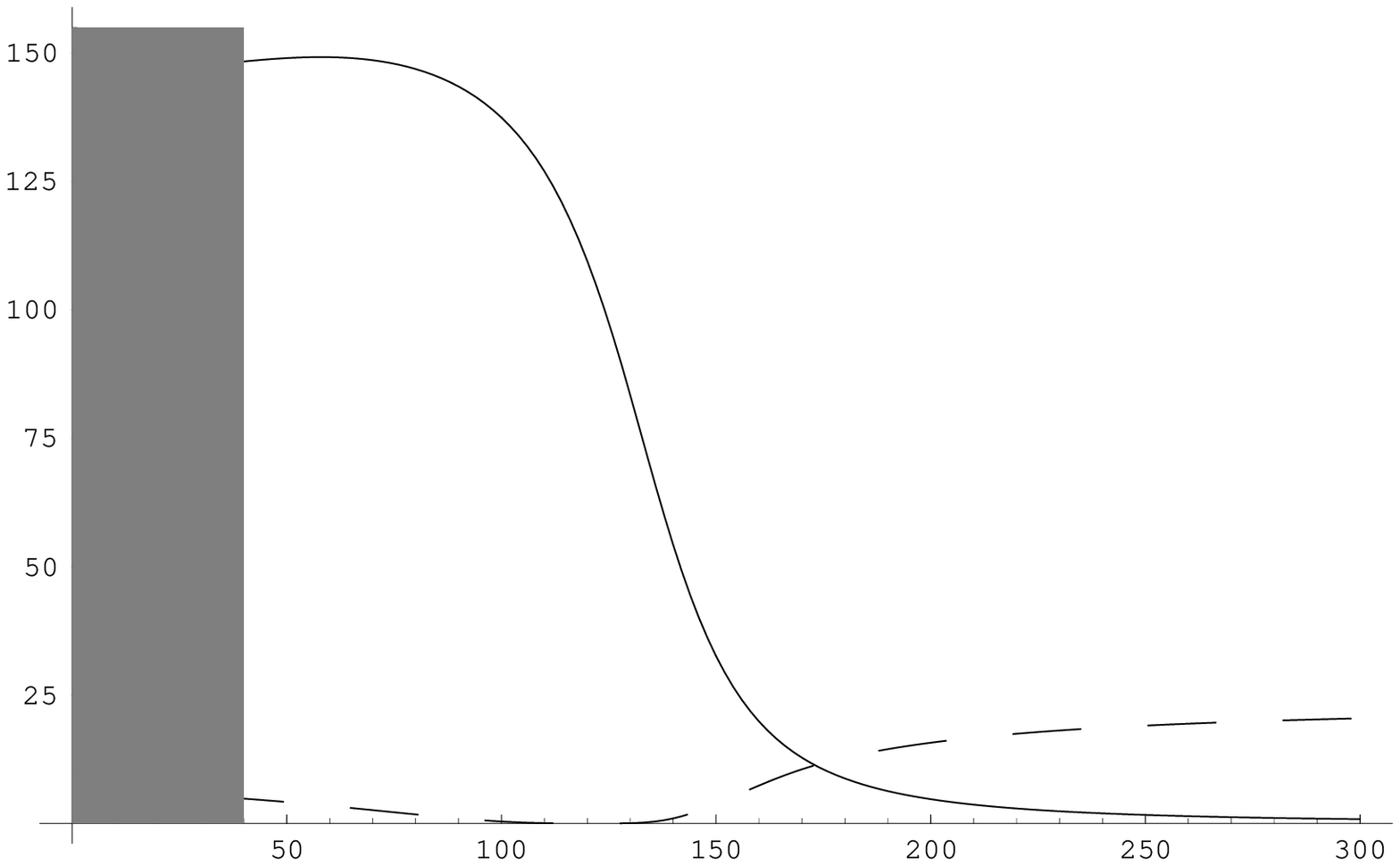}}
\put(6.0,5.25){{\tiny $M_1$/GeV}}
\put(-0.4,9.7){{\tiny $\sigma_{ee}^{L/R}$/fb}}
\put(2.5,5.2){{\tiny (a)}}
\put(6.0,1.9){\includegraphics{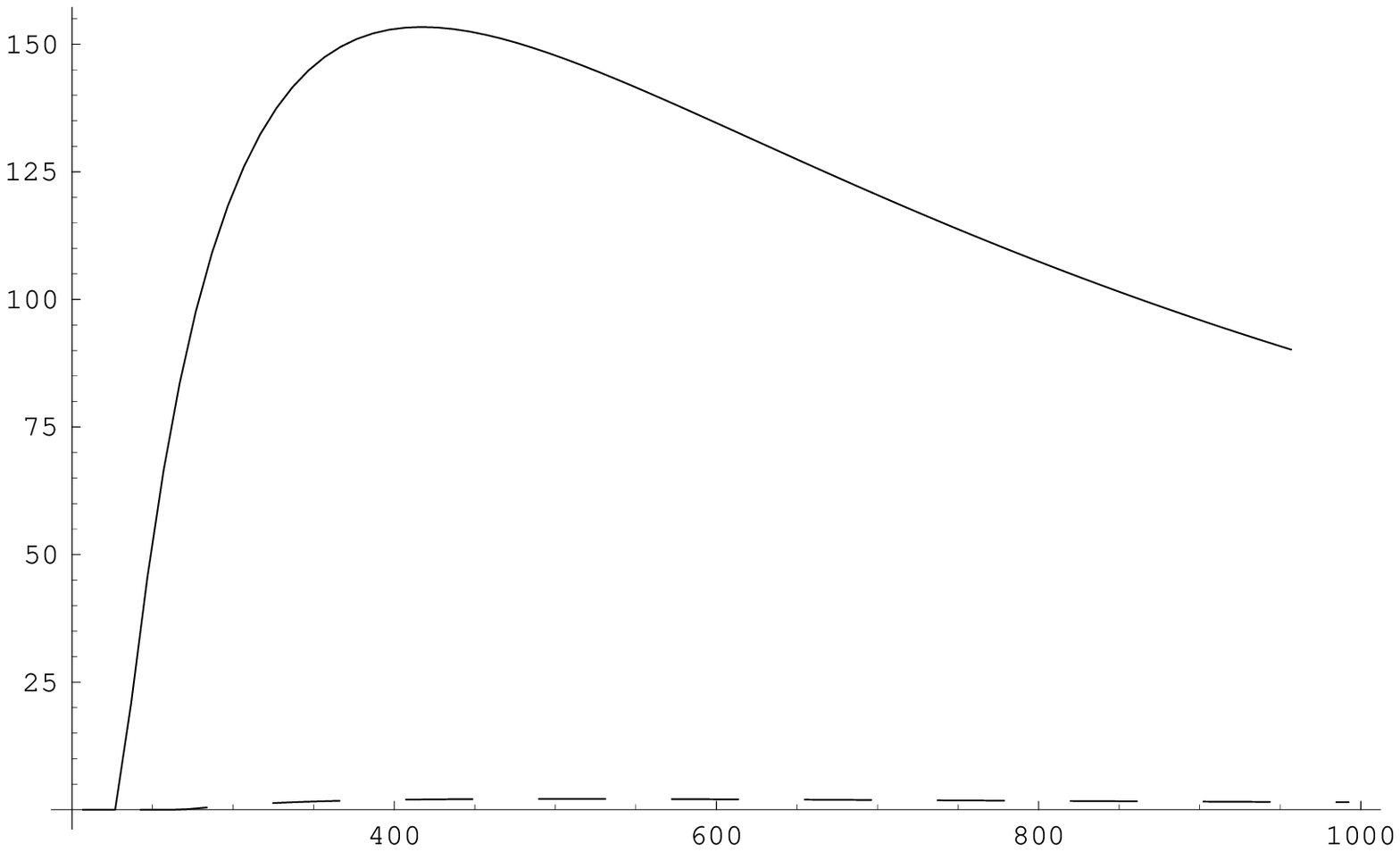}}
\put(13.2,5.3){{\tiny $\sqrt{s_{ee}}$/GeV}}
\put(7.0,9.7){{\tiny $\sigma_{ee}^{L/R}$/fb}}
\put(9.6,7.0){{\tiny $M_1=78.7$ GeV}}
\put(9.9,5.1){{\tiny (b)}}
\put(-1.4,-3.3){\includegraphics{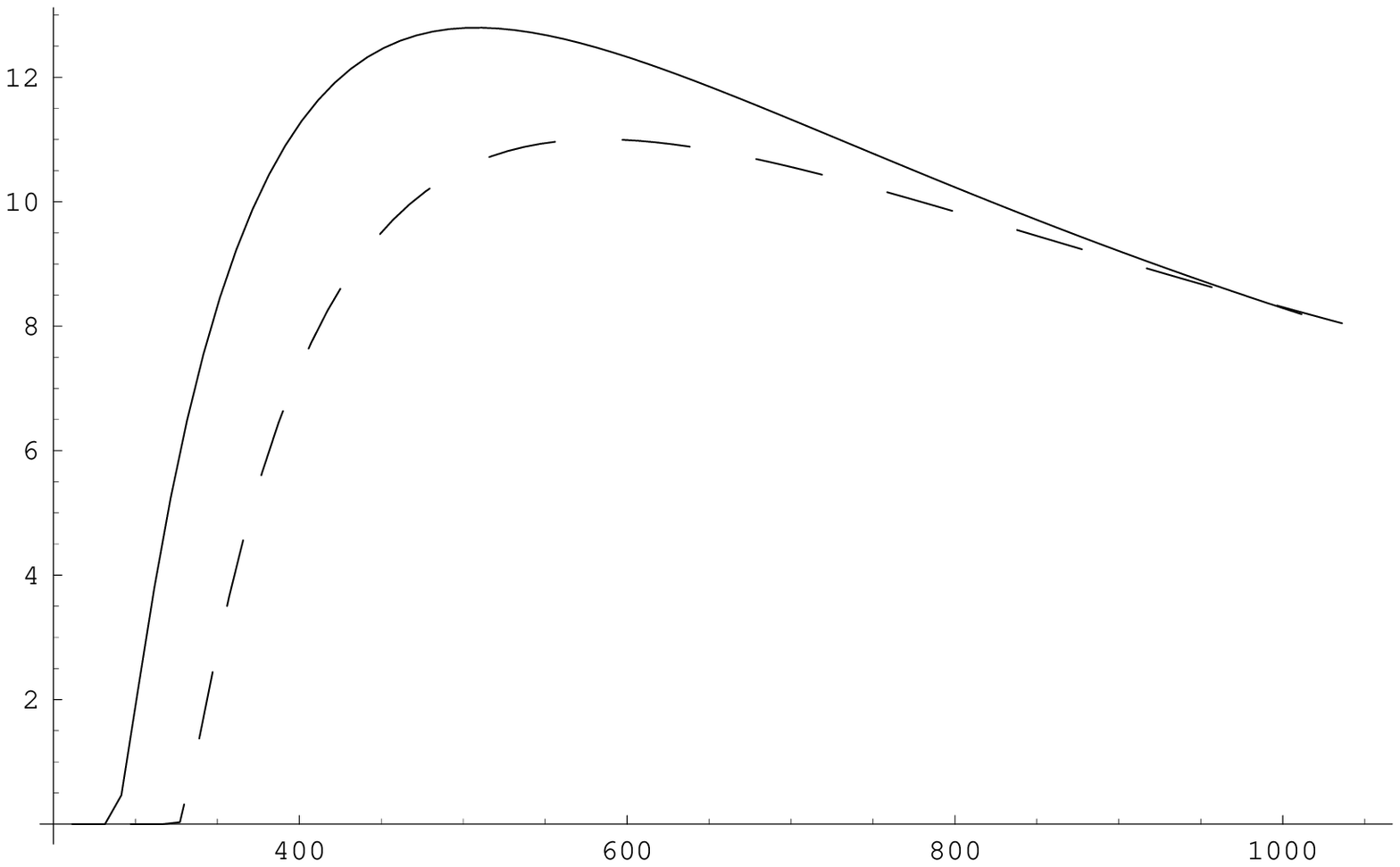}}
\put(6.0,0.1){{\tiny $\sqrt{s_{ee}}$/GeV}}
\put(-0.55,4.5){{\tiny $\sigma_{ee}^{L/R}$/fb}}
\put(2.2,1.8){{\tiny $M_1=170$ GeV}}
\put(2.5,0.0){{\tiny (c)}}
\put(6.0,-3.3){\includegraphics{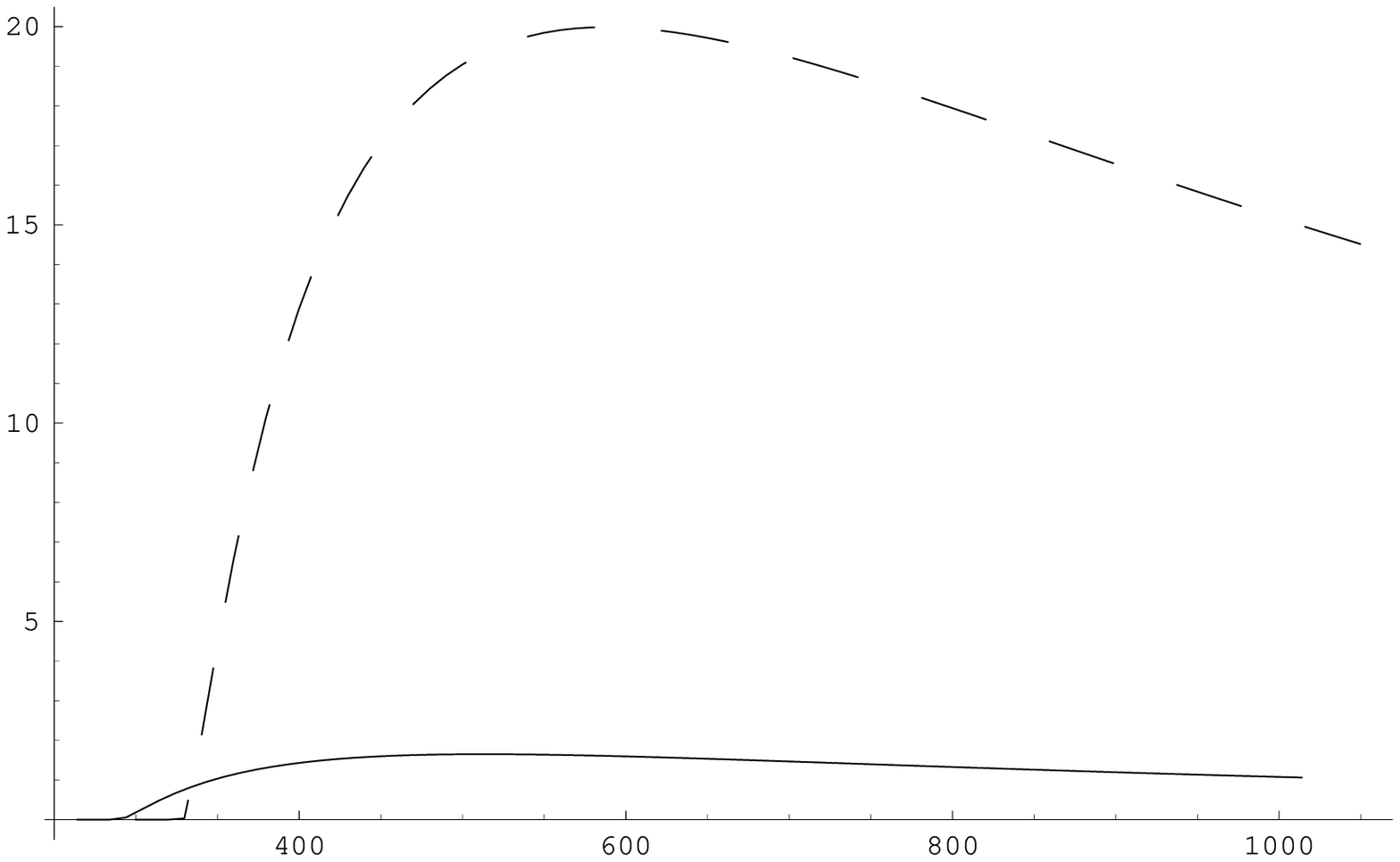}}
\put(13.2,0.1){{\tiny $\sqrt{s_{ee}}$/GeV}}
\put(7.0,4.5){{\tiny $\sigma_{ee}^{L/R}$/fb}}
\put(9.6,1.8){{\tiny $M_1=250$ GeV}}
\put(9.7,0.0){{\tiny (d)}}
\end{picture}
\caption{Total cross sections $\sigma_{ee}^R$ (solid lines) and $\sigma_{ee}^L$
(dashed lines) for $m_{\tilde{e}_R}=137.7$ GeV, $m_{\tilde{e}_L}=179.3$ GeV
and unpolarized electron and photon beams ($P_e=\lambda_e=\lambda_L=0$);
(a) $M_1$-dependence of $\sigma^{L/R}_{ee}$ for $\sqrt{s_{ee}}=500$ GeV; 
Energy dependence of $\sigma^{L/R}_{ee}$ for (b) $M_1=78.7$ GeV,  
(c) $M_1=170$ GeV and 
(d) $M_1=250$ GeV.}
\end{figure}
 
As a consequence of the somewhat higher mass the cross section for production
and decay of $\tilde{e}_L$ is additionally suppressed compared to that for
$\tilde{e}_R$. Therefore in fig.\ 2a the crossing of the cross sections is 
at a somewhat
higher value of $M_1\sim 175$ GeV than that of the couplings at $M_1\sim 150$
GeV in fig.\ 1c. For $M_1<175$ GeV the production of $\tilde{e}_R$ dominates whereas for
$M_1>175$ GeV that of $\tilde{e}_L$ dominates with, however, much smaller 
cross sections.
Fig.\ 2a  shows the strong variation of the cross section 
$\sigma_{ee}^R$ with $M_1$. If we assume that a cross section 
$\sigma_{ee}^R=100$ fb has been measured with an error
of $\pm5\%$ this is compatible
with $M_1$ between 122 GeV and 126 GeV. 

For an unpolarized electron beam ($P_e=0$) polarization of the laser beam and of the 
converted electrons essentially 
changes only the magnitude of 
the cross sections by a maximal factor between 0.7 and 1.3. As we have checked
numerically  the 
$M_1$ dependence is very 
similar to that given in fig.\ 2a.

Fig.\ 2b - 2d exhibit the energy dependence of the total cross section for
three different values of $M_1$: the GUT value $M_1=78.7$ GeV (fig.\ 2b) and two
higher values $M_1=170$ GeV (fig.\ 2c) and $M_1=250$ GeV (fig.\ 2d). 
For a polarization of the electron beam $P_e=+0.9$ ($P_e=-0.9$) 
the cross section for production and decay of left (right) 
selectrons is reduced and that for right (left) selectrons is enhanced. 

\begin{figure}[htb]
\label{asy}
\centering
\begin{picture}(14,15.5)
\put(2.3,7.2){\includegraphics{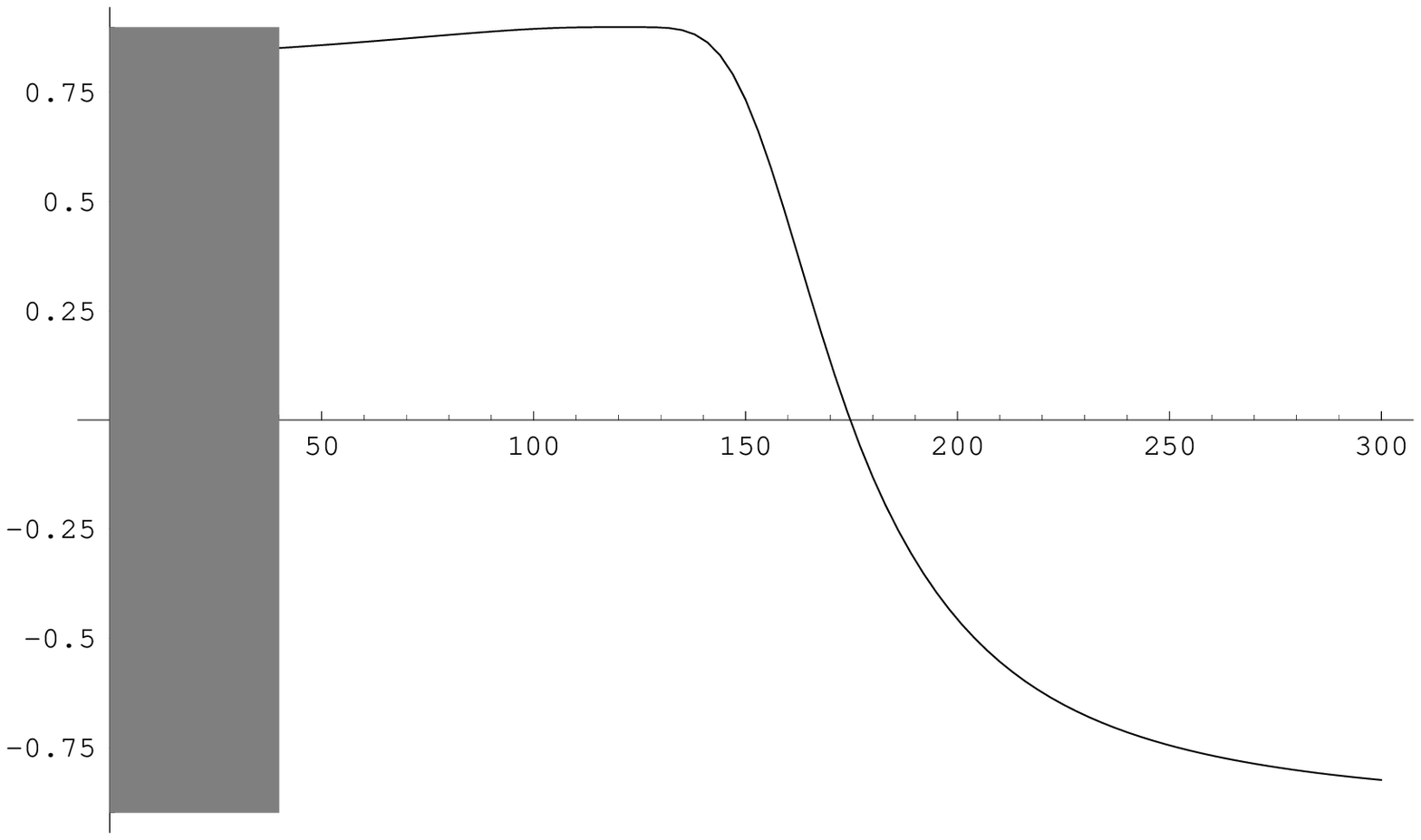}}
\put(10.0,12.5){{\tiny $M_1$/GeV}}
\put(3.7,14.9){{\tiny $A_{P_e}$}}
\put(6.2,10.5){{\tiny (a)}}
\put(-1.4,1.9){\includegraphics{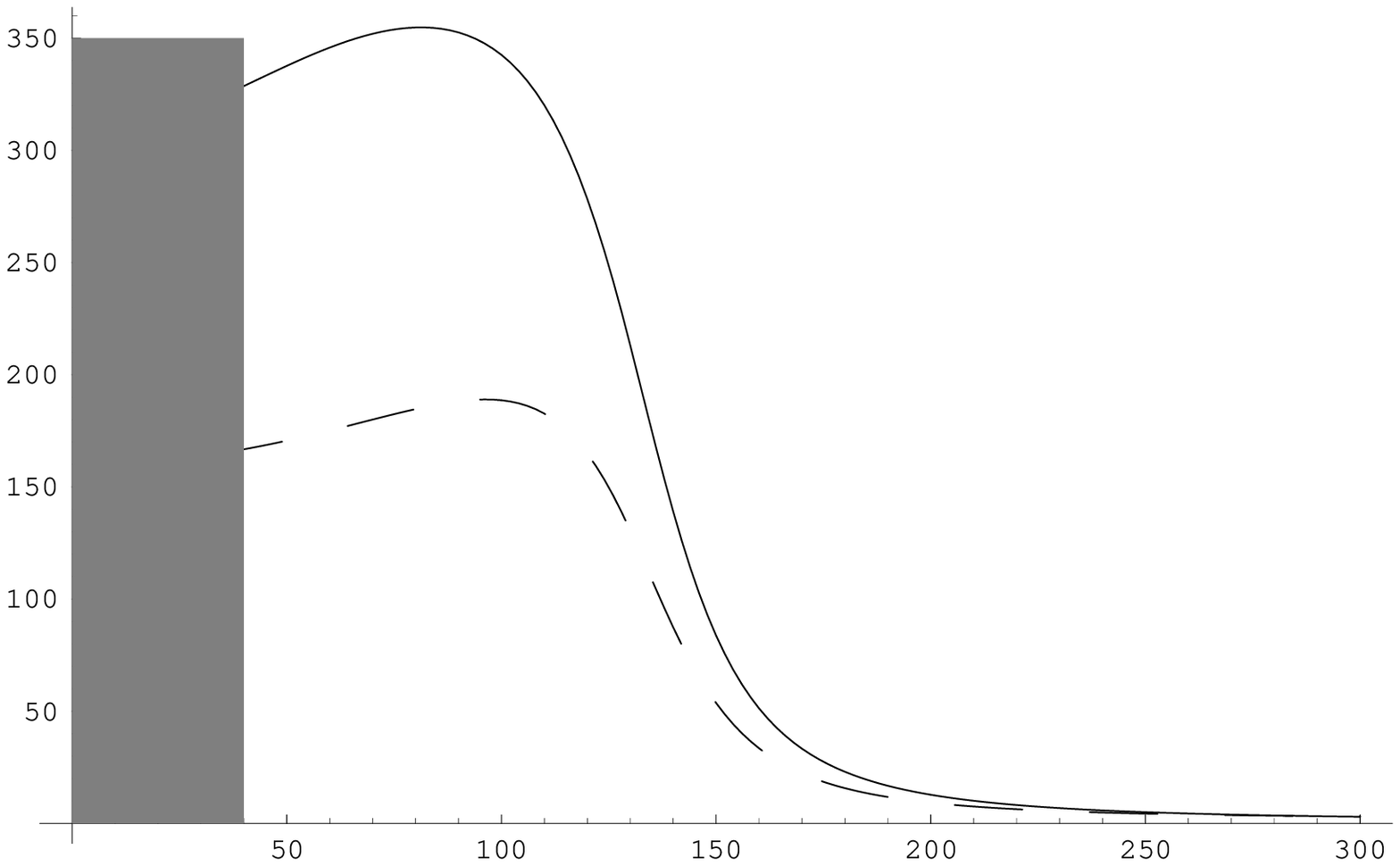}}
\put(6.0,5.2){{\tiny $M_1$/GeV}}
\put(-0.4,9.7){{\tiny $\sigma_{ee}$/fb}}
\put(2.9,5.2){{\tiny (b)}}
\put(6.0,1.9){\includegraphics{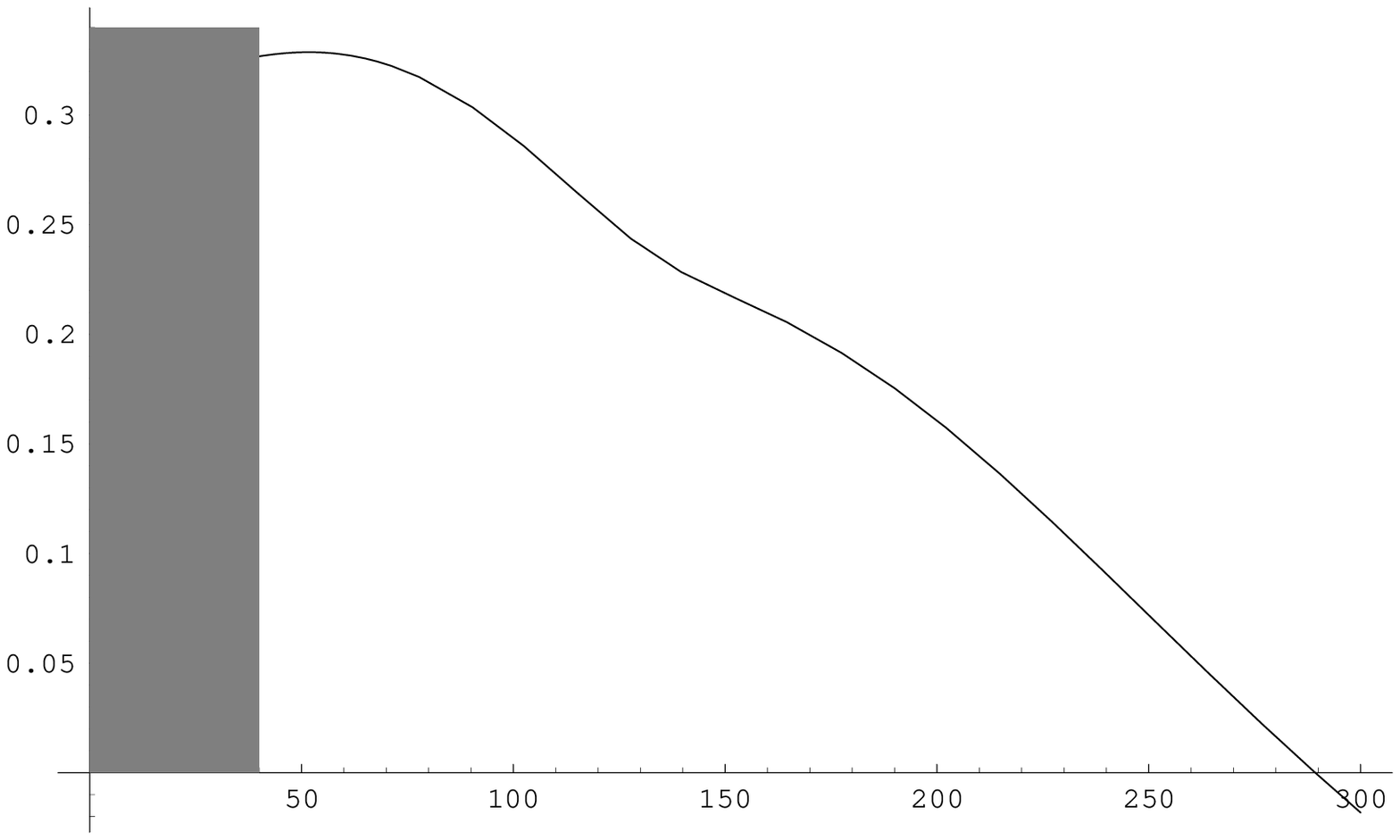}}
\put(13.4,5.4){{\tiny $M_1$/GeV}}
\put(7.0,9.6){{\tiny $A_{\lambda_L}$}}
\put(9.9,5.1){{\tiny (c)}}
\put(-1.4,-3.1){\includegraphics{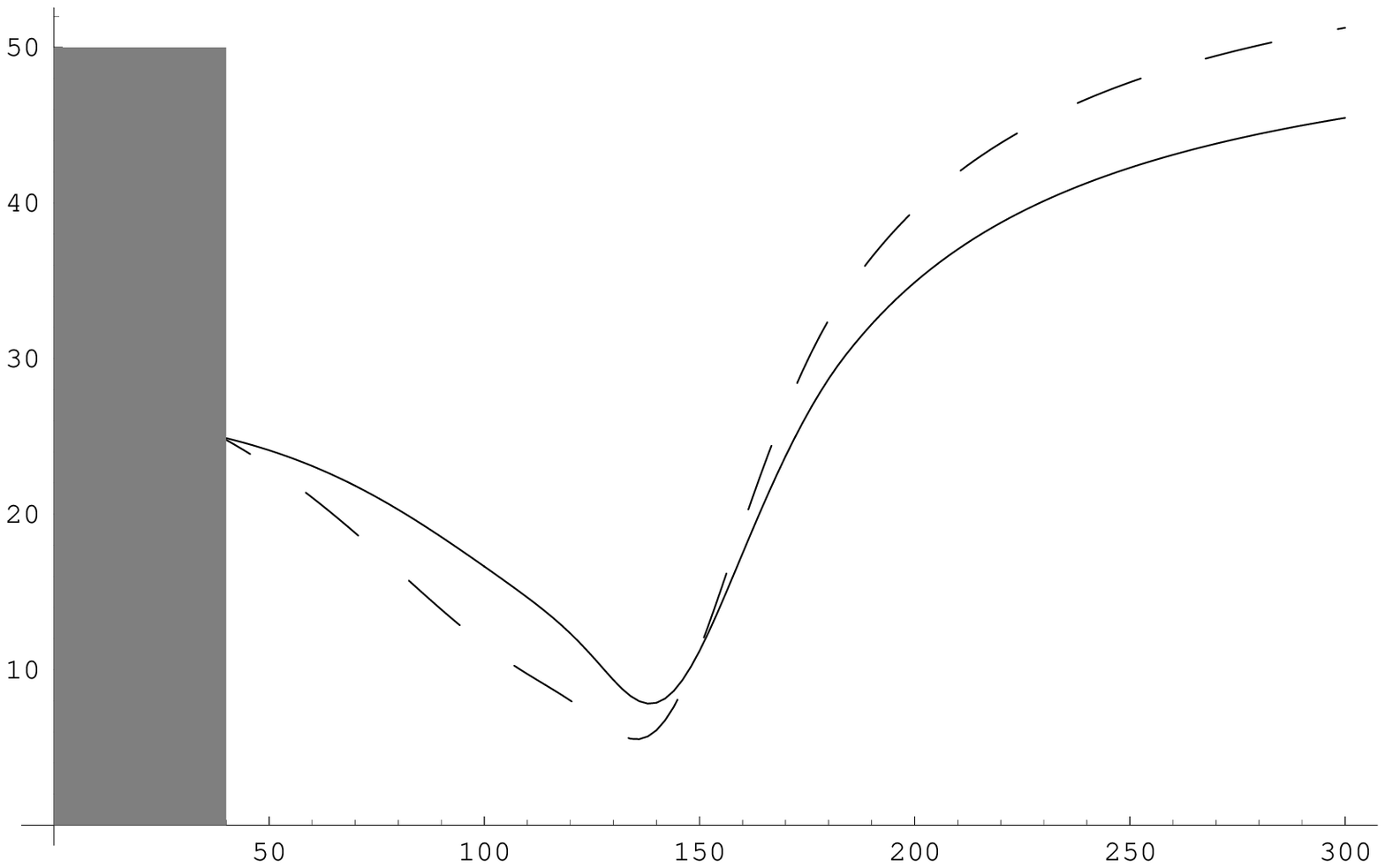}}
\put(6.0,0.2){{\tiny $M_1$/GeV}}
\put(-0.4,4.7){{\tiny $\sigma_{ee}$/fb}}
\put(2.5,0.2){{\tiny (d)}}
\put(6.0,-3.1){\includegraphics{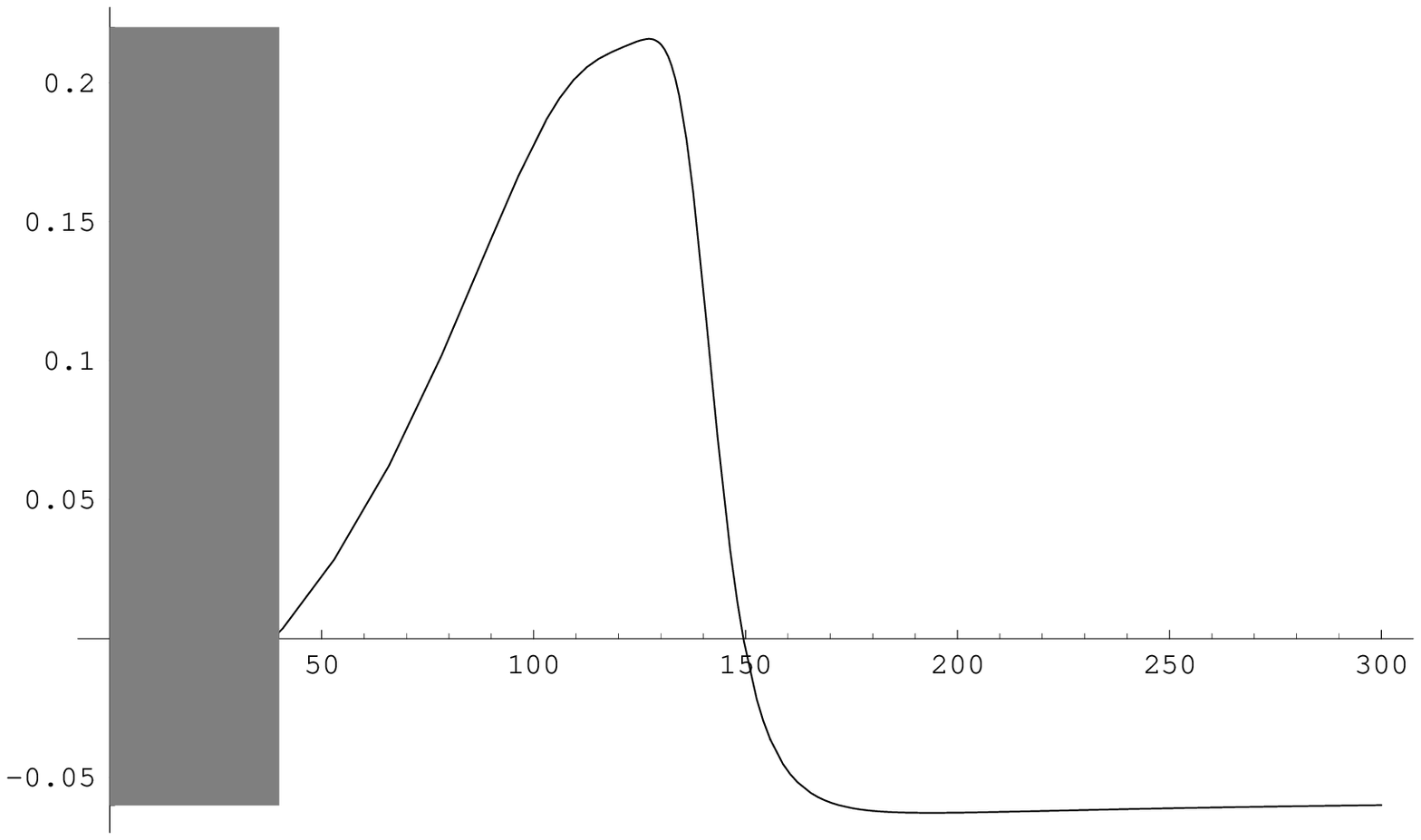}}
\put(13.4,1.1){{\tiny $M_1$/GeV}}
\put(7.0,4.6){{\tiny $A_{\lambda_L}$}}
\put(9.9,0.1){{\tiny (e)}}
\end{picture}
\caption{Total cross section $\sigma_{ee}=\sigma^L_{ee}+\sigma^R_{ee}$ and 
polarization  asymmetries for $m_{\tilde{e}_R}=137.7$ GeV and
$m_{\tilde{e}_L}=179.3$ GeV; (a) $M_1$-dependence of the asymmetry $A_{P_e}$ for  
$P_e=\pm 0.9$ and $\lambda_e= \lambda_L=0$; (b) $M_1$-dependence of  
$\sigma_{ee}$  for $P_e=0.9$, $\lambda_e=1$, $\lambda_L=+1$ (solid line)
 and for $P_e=0.9$, $\lambda_e=1$, $\lambda_L=-1$ (dashed line); 
(c) $M_1$-dependence of the asymmetry $A_{\lambda_L}$ for  
$P_e=0.9$, $\lambda_e=+1$ and $\lambda_L=\pm 1$; 
(d) $M_1$-dependence of  
$\sigma_{ee}$  for  $P_e=-0.9$, $\lambda_e=-1$, $\lambda_L=+1$ (solid line)
and for $P_e=-0.9$, $\lambda_e=-1$, $\lambda_L=-1$ (dashed line); 
(e) $M_1$-dependence of the asymmetry $A_{\lambda_L}$ for  
$P_e=-0.9$, $\lambda_e=-1$ and $\lambda_L=\pm 1$.}
\end{figure}

In fig.\ 3a the asymmetry $A_{P_e}$ defined in eq.\ (\ref{Ae2}) is shown for
unpolarized converted electrons ($\lambda_e=0$), unpolarized laser photons 
($\lambda_L=0$) and electron polarization $P_e=\pm 0.9$. In our scenario the dependence 
of $A_{P_e}$ on $\lambda_L$ and on $\lambda_e$ turns out to be negligible. The 
$M_1$-dependence of $A_{P_e}$ is as expected from that of the 
cross sections (fig.\ 2). Since for $M_1<175$ GeV ($M_1>175$ GeV) the 
production of $\tilde{e}_R$ ($\tilde{e}_L$) dominates we obtain large
positive asymmetries (large negative asymmetries) for $M_1<175$ GeV 
($M_1>175$ GeV). 
For $40\mbox{ GeV}<M_1<142$ GeV the asymmetry $A_{P_e}$ is larger than 0.85 and
nearly independent of $M_1$. In this region, however, the LSP mass (fig.\ 1a)
and the total cross section (fig.\ 2)  depend strongly on $M_1$.
For $M_1> 205$ GeV the asymmetry increases up to large negative values between 
$A_{P_e}=-0.5$ for $M_1=205$ GeV and $A_{P_e}=-0.82$ for $M_1=300$ GeV with,
however, rather small cross sections $<38$ fb. 
For $142\mbox{ GeV}<M_1<205$ GeV 
the asymmetry $A_{P_e}$ shows a strong variation with $M_1$.
If we assume that for instance an asymmetry $A_{P_e}=0.5\pm 5\%$ has been 
measured this 
is compatible with $M_1$ in the narrow region between 158 GeV and 160 GeV.

Additional informations on the value of $M_1$ can be obtained 
if the laser beam and the converted electrons are polarized. 
In fig.\ 3b we show the $M_1$-dependence of the total cross section 
$\sigma_{ee}$ for $P_e=0.9$ and  $\lambda_e=+1$. For $\lambda_L=-1$ 
ambiguities exist in the region $40\mbox{ GeV}<M_1<120$ GeV and for $M_1>180$ GeV 
the dependence on $M_1$ is rather weak. 
For $120\mbox{ GeV}<M_1<180$ GeV however this cross section shows a strong
variation with $M_1$.
For $\lambda_L=+1$ the cross section again shows ambiguities 
in the region $40\mbox{ GeV}<M_1<108$ GeV and is nearly independent on $M_1$ 
for $M_1>180$ GeV. The interval
$108\mbox{ GeV}<M_1<180$ GeV, where
the cross section is sensitive to $M_1$ is however larger than for
$\lambda_L=-1$. If we assume that a cross section
$\sigma_{ee}=250\mbox{ fb}\pm 5\%$ has been measured this is compatible with
$M_1$ between 122 GeV and 127 GeV.
In the  region $60 \mbox{ GeV}<M_1<300$ GeV the asymmetry
$A_{\lambda_L}$ (eq.\ (\ref{AL})) depicted in fig.\ 3c for $P_e=0.9$ and 
$\lambda_e=+1$
is nearly linearly dependent on $M_1$  so that it should be possible to 
determine $M_1$ uniquely in the region
$60\mbox{ GeV}<M_1<190$ GeV. 
An asymmetry $A_{\lambda_L}=0.25\pm 5\%$ would be
compatible with $M_1$ between 116 GeV and 132 GeV according to 
fig.\ 3c. In the region $M_1>190$ GeV the cross sections are 
smaller than 16 fb.
 
The cross section $\sigma_{ee}$ and the asymmetry  $A_{\lambda_L}$ are 
depicted in fig.\ 3d, e for the polarization configuration $P_e=-0.9$ and 
$\lambda_e=-1$. For $\lambda_L=-1$ the total cross section has ambiguities
in the region $40\mbox{ GeV}<M_1<167$ GeV and for $\lambda_L=+1$ in
the region $40\mbox{ GeV}<M_1<173$ GeV. For $M_1>173$ GeV one notices a strong
variation of the 
cross section for $\lambda_L=\pm1$. As can be seen from fig.\ 3d with
$\lambda_L=+1$
a cross section $\sigma_{ee}=35\mbox{ fb}\pm5\%$ 
is compatible with $M_1$ between 193 GeV and 209 GeV. 
For this polarization configuration the asymmetry $A_{\lambda_L}$ (fig.\ 3e) 
grows practically 
linearly between $M_1=40$ GeV and $M_1=126$ GeV 
 and is very sensitive on $M_1$ but shows ambiguities between $M_1=40$ GeV and $M_1=150$ GeV . If we assume that an asymmetry 
$A_{\lambda_L}=0.15\pm5\%$ has been measured this is compatible with $M_1$ 
between 89 GeV and 94 GeV or between 138 GeV and 140 GeV according to fig.\ 3e.
 One can distinguish between these two regions via the cross section for 
$\lambda_L=+1$ depicted in fig.\ 3d because one expects 18-19 fb for $M_1$ 
between 89 GeV and 94 GeV and 7-8 fb for $M_1$ between 138 GeV and 140 GeV.
For $M_1>170$ GeV the 
asymmetry is nearly constant
$A_{\lambda_L}\sim -0.07$.  

To sum up: for unpolarized laser beams ($\lambda_L=0$) and converted 
electrons ($\lambda_e=0$) the polarization asymmetry 
$A_{P_e}$ exhibits a pronounced $M_1$ dependence in the region
$142\mbox{ GeV}<M_1<205$ GeV. For the polarization configuration
$P_e=0.9$, $\lambda_e=+1$ and $\lambda_L=\pm 1$ the cross sections 
$\sigma_{ee}$ and the polarization asymmetry  $A_{\lambda_L}$
are sensitive to $M_1$ in the region $60\mbox{ GeV}<M_1<190$ GeV. 
Finally  for $P_e=-0.9$, $\lambda_e=-1$ and 
$\lambda_L=\pm 1$ these observables show a strong $M_1$ dependence in the 
region $40\mbox{ GeV}<M_1<300$ GeV.

\begin{figure}[htb]
\label{asy2}
\centering
\begin{picture}(14,5.5)
\put(-1.4,-3.1){\includegraphics{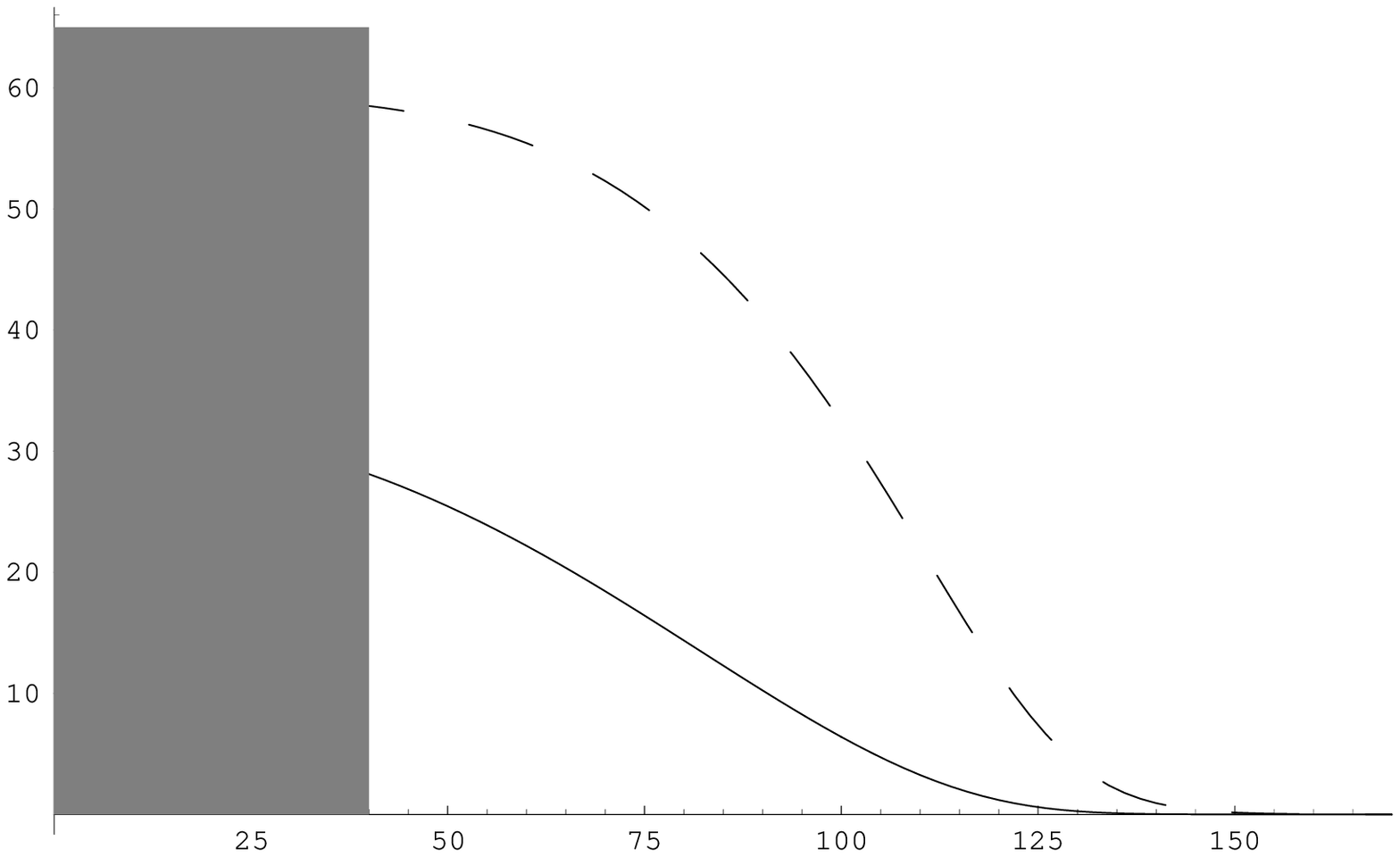}}
\put(6.0,0.4){{\tiny $M_1$/GeV}}
\put(-0.4,4.7){{\tiny $\sigma_{ee}$/fb}}
\put(2.5,0.2){{\tiny (a)}}
\put(6.0,-3.1){\includegraphics{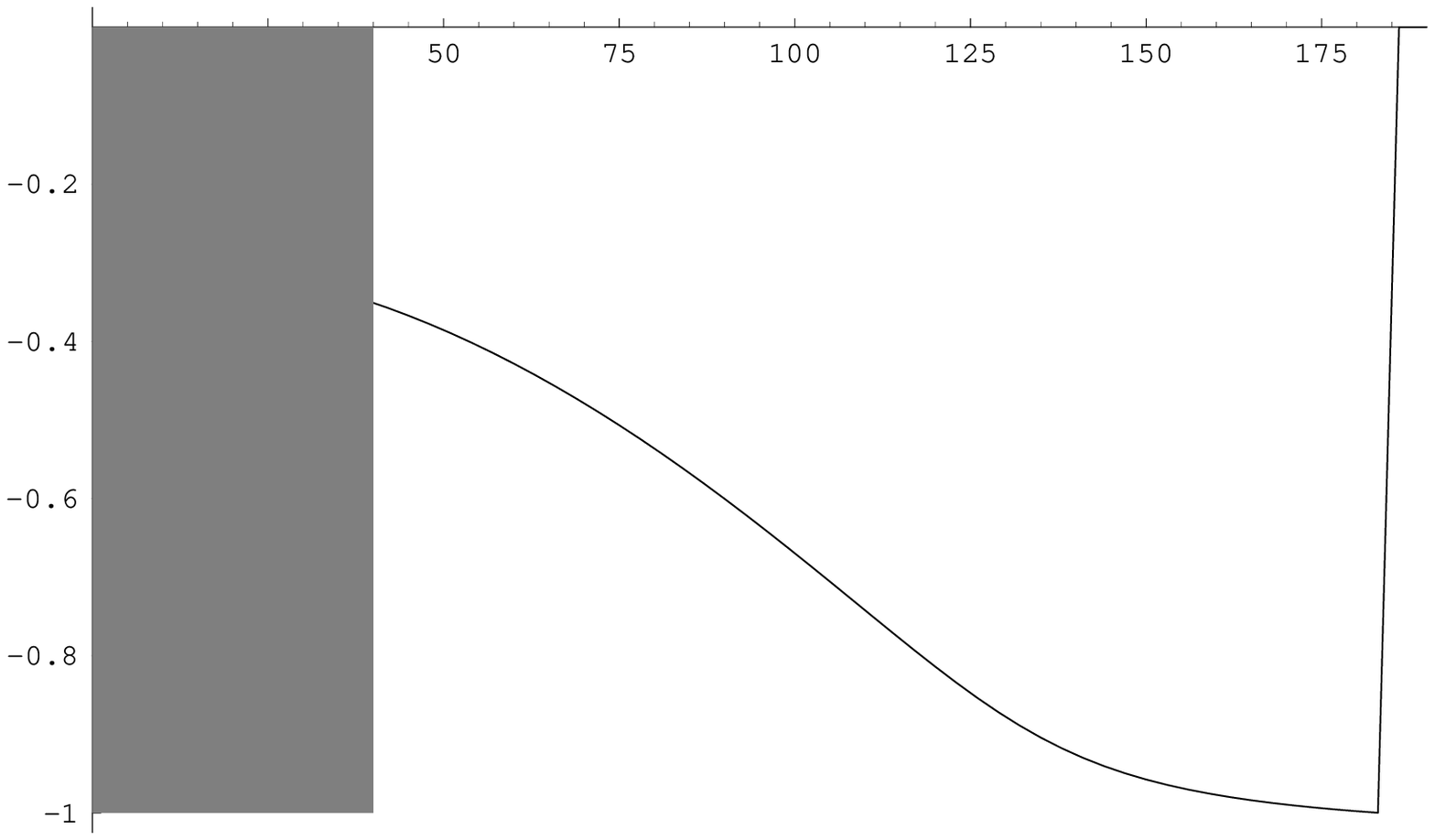}}
\put(13.4,4.5){{\tiny $M_1$/GeV}}
\put(7.0,4.6){{\tiny $A_{\lambda_L}$}}
\put(9.9,0.1){{\tiny (b)}}
\end{picture}
\caption{Total cross section $\sigma_{ee}=\sigma^L_{ee}+\sigma^R_{ee}$ and polarization
asymmetry $A_{\lambda_L}$ for $m_{\tilde{e}_R}=330.5$ GeV and 
$m_{\tilde{e}_L}=350.0$ GeV; (a) $M_1$-dependence of  $\sigma_{ee}$
for $P_e=0.9$, $\lambda_e=1$, $\lambda_L=+1$ 
(solid line) and $P_e=0.9$, $\lambda_e=1$, $\lambda_L=-1$ (dashed line);
(b) $M_1$- dependence of $A_{\lambda_L}$ 
for $P_e=0.9$, $\lambda_e=+1$ 
and $\lambda_L=\pm 1$.}
\end{figure}

We choose as a second example higher selectron masses 
$m_{\tilde{e}_L}=350.0$ GeV and $m_{\tilde{e}_R}=330.5$ GeV corresponding to 
 $m_0=320$ GeV.
Then for $\sqrt{s_{ee}}=500$ GeV selectron pair production in $e^+e^-$
annihilation is forbidden, whereas single selectron production in 
$e^-\gamma \longrightarrow \tilde{\chi}_1^0\tilde{e}_{L/R}^-$ is still
possible, provided that 
$\sqrt{s_{e\gamma}}>m_{\tilde{e}_{L/R}^-}+m_{\tilde{\chi}_1^0}$ where
$\sqrt{s_{e\gamma}} \sim 0.91\cdot\sqrt{s_{ee}}$ is the energy of the hardest photon 
obtained by Compton backscattering \cite{stefan}.
Now the kinematical accessible $M_1$ region is confined to $M_1<184$ GeV 
($m_{\tilde{\chi}_1^0}<124.6$ GeV).
In fig.\ 4a,b we show the total cross section and the asymmetry  $A_{\lambda_L}$
for $P_e=0.9$, $\lambda_e=+1$ and $\lambda_L=\pm 1$. For $\lambda_L=+1$ the 
cross section depends nearly linearly on $M_1$ in the region 
$40\mbox{ GeV}<M_1<115$ GeV. For $M_1>115$ GeV the cross section is smaller than 
2 fb. The cross section for $\lambda_L=-1$ is higher and more sensitive to
$M_1$ between 
$40\mbox{ GeV}<M_1<135$ GeV. If we assume for example that a cross section 
$\sigma_{ee}=45\mbox{ fb}\pm5\%$ has been measured this is compatible with 
 $M_1$ between 
80 GeV and 88 GeV. Also the polarization asymmetry $A_{\lambda_L}$ strongly
depends on $M_1$ in the whole region. According to fig.\ 4b an asymmetry 
$A_{\lambda_L}= -0.7\pm5\%$  would be 
compatible with  $M_1$ between 99 GeV and 109 GeV. 
The polarization asymmetry $A_{P_e}$  for this scenario is between 0.85
and 0.9 and depends only weakly on $M_1$. Also the polarization
configuration $P_e=-0.9$, $\lambda_e=-1$ and $\lambda_L=\pm 1$ is not
shown because the cross sections are smaller than 2 fb.
Thus for the case of high selectron masses and polarization configuration  
$P_e=0.9$, $\lambda_e=+1$ and $\lambda_L=\pm 1$ both the cross section 
and the asymmetry  $A_{\lambda_L}$ can be helpful for determining
 $M_1$ in the greatest part ($40\mbox{ GeV}<M_1<135$ GeV) of the 
kinematical accessible region $M_1<184$ GeV.

\section{Conclusion}
We have demonstrated that associated selectron - LSP production with subsequent
leptonic decay of the selectron
$e^-\gamma \longrightarrow \tilde{\chi}_1^0\tilde{e}_{L/R}^-
\longrightarrow e^-\tilde{\chi}_1^0\tilde{\chi}_1^0$ at a $\sqrt{s_{ee}}=500$
GeV linear collider in the $e\gamma$ mode should allow to test for a 
gaugino-like LSP the GUT relation $M_1=M_2\cdot\frac{5}{3}\tan^2\theta_W$ 
between the MSSM gaugino mass parameters.
The polarization $P_e$ of the electron beam 
 helps to enlarge the production cross section for left or right selectrons. 
For suitably polarized electron beams and laser photons the total cross section
$\sigma_{ee}$ and the polarization asymmetries $A_{P_e}$ and $A_{\lambda_L}$
are very sensitive to the gaugino mass parameter $M_1$ in the whole 
investigated region between 40 GeV and 300 GeV. For high selectron masses
$m_{\tilde{e}_{L/R}}$ the accessible $M_1$ region is kinematically constrained. The
optimal polarization configuration depends on the values of the selectron 
masses. For realistic predictions a complete MC study with inclusion of 
background processes and experimental cuts would be indispensable.

\section{Acknowledgements}
We are grateful to Gudrid Moortgat-Pick and Stefan Hesselbach for valuable
discussions.
This work was supported by the Deutsche Forschungsgemeinschaft under contract 
no. FR 1064/4-1 and the Bundesministerium f\"ur Bildung und Forschung (BMBF) 
under contract number 05 HT9WWA 9.

\end{document}